\documentclass[12pt,preprint]{aastex}

\slugcomment{}

\shorttitle{GRB\,071112C}
\shortauthors{Huang et al.}

\begin{document}

\title{THE GRB\,071112C: A CASE STUDY OF DIFFERENT MECHANISMS IN X-RAY AND OPTICAL TEMPORAL EVOLUTION}
\author{
K. Y. \textsc{Huang}\altaffilmark{1},
Y. \textsc{Urata}\altaffilmark{2},
Y. H. \textsc{Tung}\altaffilmark{1,3},
H. M. \textsc{Lin}\altaffilmark{1,2},
L. P. \textsc{Xin}\altaffilmark{4},
M. \textsc{Yoshida}\altaffilmark{5,6},
W. \textsc{Zheng}\altaffilmark{7},
C. \textsc{Akerlof}\altaffilmark{7},  
S. Y. \textsc{Wang}\altaffilmark{1},
W.  H. \textsc{Ip}\altaffilmark{2}, 
M. J. \textsc{Lehner}\altaffilmark{1,8,9},
F. B. \textsc{Bianco}\altaffilmark{10,11},  
N. \textsc{Kawai}\altaffilmark{12},
D. \textsc{Kuroda}\altaffilmark{6},
S. L. \textsc{Marshall}\altaffilmark{13},
M. E. \textsc{Schwamb}\altaffilmark{14},
Y. \textsc{Qiu}\altaffilmark{4},
J. H. \textsc{Wang}\altaffilmark{1}, 
C. Y. \textsc{Wen}\altaffilmark{1}, 
J. \textsc{Wei}\altaffilmark{4},
K. \textsc{Yanagisawa}\altaffilmark{6},
and Z. W. \textsc{Zhang}\altaffilmark{1}
}

\altaffiltext{1}{Institute of Astronomy and Astrophysics, Academia Sinica, P.O. Box 23-141, Taipei 106, Taiwan}
\altaffiltext{2}{Institute of Astronomy, National Central University, Chung-Li 32054, Taiwan}
\altaffiltext{3}{Institute of Astronomy and Department of Physics, National Tsing Hua University, Hsinchu 30013, Taiwan}
\altaffiltext{4}{National Astronomical Observatories, Chinese Academy of Sciences, Beijing, 100012, China}
\altaffiltext{5}{Hiroshima Astrophysical Science Center, Hiroshima University, 1-3-1 Kagamiyama, Higashi-Hiroshima, Hiroshima 739-8526, Japan}
\altaffiltext{6}{Okayama Astrophysical Observatory, National Astronomical Observatory of Japan, Kamogata, Asakuchi, Okayama 719-0232, JApan}
\altaffiltext{7}{Department of Physics, University of Michigan, Ann Arbor, MI 48109, USA}
\altaffiltext{8}{Department of Physics and Astronomy, University of Pennsylvania, 209 South 33rd Street, Philadelphia, PA 19104, USA}
\altaffiltext{9}{Harvard-Smithsonian Center for Astrophysics, 60 Garden Street, Cambridge, MA 02138, USA}
\altaffiltext{10}{Department of Physics, University of California Santa Barbara, Mail Code 9530, Santa Barbara, CA 93106-9530, USA}
\altaffiltext{11}{Las Cumbres Observatory Global Telescope Network, Inc. 6740 Cortona Drive, Suite 102, Santa Barbara, CA 93117, USA }
\altaffiltext{12}{Department of Physics, Tokyo Institute of Technology, 2-21-1 Ookayama, Meguro-ku, Tokyo 152-8551, Japan}
\altaffiltext{13}{Kavli Institute for Particle Astrophysics and Cosmology, 2575 Sand Hill Road, MS 29, Menlo Park, CA 94025, USA}
\altaffiltext{14}{Department of Physics, Yale University, P.O. Box 208121, New Haven, CT 06520-8121, USA}


%\begin{abstract}
We present the study on GRB\,071112C X-ray and optical light curves.
In these two wavelength ranges, we have found different temporal
properties. The $R$-band light curve showed an initial rise followed
by a single power-law decay, while the X-ray light curve was described
by a single power-law decay plus a flare-like feature. Our analysis
shows that the observed temporal evolution cannot be described by the
external shock model in which the X-ray and optical emission are
produced by the same emission mechanism. No significant color changes
in multi-band light curves and a reasonable value of the initial
Lorentz factor ($\Gamma_0 = 275 \pm 20 $) in a uniform interstellar
medium support the afterglow onset scenario as the correct
interpretation for the early $R$-band rise. The result suggests that
the optical flux is dominated by afterglow. Our further investigations
show that the X-ray flux could be created by an additional feature
related to energy injection and X-ray afterglow. Different theoretical
interpretations indicate the additional feature in X-ray can be
explained by either late internal dissipation or local inverse-Compton
scattering in the external shock.

%\end{abstract}


\keywords{gamma-ray burst : individual: GRB 071112C}

\section{INTRODUCTION}

The {\it Swift} Gamma-Ray Explorer, launched in late 2004, has
provided accurate positions for many gamma-ray bursts (GRBs), enabling
a number of early X-ray and optical observations. The on-board X-Ray
Telescope (XRT) data revealed that many GRBs have complicated
evolutions (e.g., flare and shallow decay) and that their X-ray temporal
evolution could be described by a three-component generic broken power
law with an initial steep power-law decay ($F(\nu,t) \propto
t^{-3\sim-5}$), followed by a very shallow decay ($F(\nu,t) \propto
t^{-0.5}$) or a flare, finally changing to a steep decay ($F(\nu,t)
\propto t^{-0.9\sim -1.3}$; \citet{nousek06,zhang06}). These
properties could be characterized by prompt emission from GRBs or a
mixture of different emission components contributing to the observed
X-ray emission \citep{will07,zhang07,liang09,nardini10}. Early optical
afterglow behaviors also show more diverse properties compared to the
simpler late time evolutions. \citet{pan08} presented early afterglow
behaviors of 28 known-redshift GRBs and grouped those GRBs as (1)
fast-rising with peaks at about 100~s; (2) slow-rising with peaks
after 100~s, (3) fast decay and (4) plateau. They proposed that the
angular asymmetry of the GRB ejecta viewed along different lines of
sight generates the diversity of early optical afterglow light
curves. A similar study with more samples was carried out by
\citet{kann10} as well. They concluded that about 60\% of early
optical detections are the forward-shock-dominated afterglows.

 Comparisons of X-ray and optical light curve have clearly shown that
 the evolutions for the two wavelengths are generally different
 \citep{pan06,huang07a}. Significantly, there are GRBs whose decay
 indices during late-time evolution could not be explained by the
 external shock model in which X-ray and optical emission must be
 produced by the same emission mechanism \citep{urata07}. These
 studies suggest that the X-ray and optical emission are generated by
 different outflows. To explain possible emission mechanisms and
 provide reasonable interpretations, \citet{ghise07} proposed a late
 prompt emission scenario. They suggested that the late internal
 shocks with lower power could be created and superposed on the real
 afterglow emission of prompt GRB emission. However, it is still
 unclear how the early temporal evolution is affected by prompt
 emission and how long the prompt emission can sustain the overall
 temporal evolution. More extensive coverage of X-ray and optical
 observations are essential to solve this problem.

Recently, \citet{liang10} analyzed 32 GRBs with early smooth bumps in
their early optical or X-ray light curves and investigated a possible
relation with the initial Lorentz factor. They found that early bright
X-ray emissions are usually dominated by non-forward shock
components, but sometimes the forward shock emissions are observable
in the X-ray wavelength, and an achromatic feature in X-ray and
optical is observed. In the study, they also discovered a good
correlation between the initial Lorentz factor and the GRB apparent
isotropic energy. Here, we examine our optical measurements of
GRB\,071112C as well as the corresponding simultaneous X-ray
observations to explore their possible emission mechanisms.

% Introduction of GRB 071112C
On 2007 November 12, the {\it Swift} Burst Alert Telescope (BAT)
detected GRB\,071112C at 18:32:57 UT. This burst showed a single fast
rise-exponential decay peak and the measured $T_{90}$ ($15$-$350$~keV)
was $15 \pm 12$~s. The 80$\%$ total fluence in the $15$-$150$~keV band
was $(3.0 \pm 0.4)\times 10^{-6} {\rm erg}~{\rm cm^{-2}}$, which
corresponded to a lower limit of isotropic energy ${E}_{\rm iso} = 5.3
\times 10^{51}$~erg at a redshift $z= 0.823$ (assume $H_0 = 70~{\rm
  km}~{\rm s^{-1}~Mpc^{-1}}, \Omega_{\rm M} = 0.3$, and
$\Omega_{\Lambda} = 0.7$). The XRT started to observe this burst from
84~s after the BAT triggered. The XRT observations showed a smooth
re-brightening around $t=500$~s after the burst and followed a simple
decay \citep{stratta}. Two robotic optical telescopes, the ROTSE-IIIc
and the TAOS, responded to this burst at 65~s and 93~s after the
burst, respectively \citep{yuan, huang07b}. In addition, the optical
afterglow was observed by a number of ground telescopes in $V$, $R$,
$I$, $J$, and $K$ bands
\citep{uemura,wang,klotz,burenin,chen,nugent,dintinjana,ishimura,greco,stefano,yoshida,uemurab,minezaki}. The
spectral measurements of the optical afterglow by the Very Large
Telescope and the Gemini North telescope indicated that the redshift
of this burst was 0.823 \citep{jakobsson,cucchuara07}.

\section{OPTICAL AND X-RAY TEMPORAL ANALYSIS}
The ROTSE-IIIc and TAOS optical observations started around $t=60$~s.
The ROTSE-IIIc telescope \citep{akerlof03} detected the GRB\,071112C
optical afterglow with brightness $R=17.1 \pm 0.2$ at $t=90.9$~s. At
the same time, the optical afterglow was also detected by TAOS-A and
TAOS-B telescopes \citep{lehner09} with sequences of 1~s and 5~s
exposures, respectively. Subsequently, a series of optical multi-band
follow-up observations were also carried out by Xinglong 0.8~m and
1.0~m telescope in China \citep{zhangwk08}, the Lulin One-meter
telescope (LOT) in Taiwan \citep{huang05,urata05}, and the 0.5~m
MITSuME telescope in Japan \citep{kotani05}. About one year after the
burst, the host galaxy of GRB\,071112C was clearly detected with the
3.8~m Canda-Feance-Hawaii Telescope (CFHT). The log of our optical
observations is summarized in Table~1.

The optical images were processed by a standard procedure including
bias and dark subtraction and flat-fielding using IRAF. The DAOPHOT
package was used to perform point-spread function (PSF)
photometry. Seven bright stars in the images were used to create the
PSF model. The absolute photometric calibration for GRB field was
determined using Landolt \citep{landolt} field SA~92, SA~95 and
PG~0231+051 with a range of airmass. For calibration, we use 17
reference stars with colors similar to the afterglow ($V-R \sim 0.4$)
. Both photometric and systematic errors were included in the
magnitude error. Besides our own observations, we re-calibrated the
reported afterglow measurements by using the reference stars in the
GRB\,071112C field. Several measurements in GCN reports were
calibrated by USNOB stars; the USNOB stars are on average 0.05
brighter than the stars of our calibration. Since the reference stars
were provided by \citet{burenin} and \citet{uemura, uemurab}, we
measured these stars from our LOT $R$-band and Xinglong $V$-band
images and obtained their averaged magnitude and rms errors. The
results were then used to re-calibrated the reported afterglow
magnitude. The uncertainties with 1$\sigma$ level confidence were
adopted in this paper.

As shown in Figure~1, the $V$-, $R$- and $I$-band light curves of
GRB\,071112C can be expressed in terms of a power law with $F(t)
\propto t^{-\alpha_{\rm opt}}$. Here, each $\alpha_{\rm opt}$ is the
power-law index in each optical band. We find $\alpha_{\rm V} = 1.02
\pm 0.05$ ($\chi^{2}/\nu$= 1.51 for $\nu=28$) from $V$-band data,
$\alpha_{\rm R} = 0.85 \pm 0.02$ ($\chi^{2}/\nu$= 2.16 for $\nu=76$)
from $R$-band data, and $\alpha_{\rm I} = 0.96 \pm 0.05$
($\chi^{2}/\nu$= 1.79 for $\nu=5$) from $I$-band data. Besides our
optical data, we also analyzed the calibrated NIR measurements by
\citet{uemurab} and \citet{minezaki}. The $J$- and $K$-band light
curves could be expressed by a single power law with a index
$\alpha_{\rm J} = 0.99 \pm 0.04$ ($\chi^{2}/\nu$= 0.38 for $\nu=8$)
and $\alpha_{\rm K} = 0.83 \pm 0.04$ ($\chi^{2}/\nu$= 0.18 for
$\nu=3$), respectively.

Note that the $R$-band light curve demonstrates a plateau in the early
evolution. We next fit the $R$-band measurements with a broken
power-law function, $F(\nu,t) = {F_{\nu}^{\ast}} / [(t/t_{\rm
    Rb})^{\alpha_{\rm R1}}+(t/t_{\rm Rb})^{\alpha_{\rm R2}}]$, where
$t_{\rm Rb}$ is the break time in $R$-band light curve, $\alpha_{\rm
  R1}$ and $\alpha_{\rm R2}$ are the power-law indices before and
after the $R$-band break time $t_{\rm Rb}$, and $F_{\nu}^{\ast}$ is
the flux at $t_{\rm Rb}$.  We obtain $\alpha_{\rm R1} = -1.54 \pm
0.62$, $\alpha_{\rm R2} = 0.92 \pm 0.02$, and $t_{\rm Rb} = 99.4 \pm
7.3$~s ($\chi^{2}/\nu$= 1.25 for $\nu=74$). The smaller chi-square value
indicates that the broken power-law function is a reasonable fitting
function to explain the $R$-band evolution in GRB\,071112C and implies
a rising behavior in early $R$-band light curve.

Figure~2 shows the $R$-band and XRT light curves of GRB\,071112C. The
XRT $0.3$-$10$\,keV light curve was downloaded from the {\it
  Swift}/XRT GRB light curve repository \citep{evans}.  To plot the
$R$-band and X-ray light curves on a consistent scale, we converted
the afterglow brightness to units of mJy. It is clear that a
Gaussian-shaped flare appeared in the XRT light curve around $t=500$~s
after the burst.  The XRT light curve can be fit with a single power
law plus a Gaussian function ($F(t) = A_0 \times t^{-\alpha_{\rm
    x}}+A_1\times e^{-(t-A_2)^2/2A_3^2}$), where $A_0$ is a constant
value, $\alpha_{\rm x}$ is the single power-law index, $A_1$ is peak
intensity at peak position $A_2$,and $A_3$ is the width of the
Gaussian feature.  The best-fit parameters are $\alpha_{\rm x} = 1.36
\pm 0.02$, $A_0 = 23.51 \pm 2.12$, $A_1 = 4.23 \pm 0.5\,{\rm \mu Jy}$,
$A_2 = 763.13 \pm 35.05$\,s, and $A_3 = 274.91 \pm 33.44$\,s
($\chi^{2}/\nu$= 1.27 for $\nu =117$). If we exclude the flare
component, the overall XRT light curve could be well fit by a single
power law with an index $\alpha_{\rm x} = 1.36 \pm 0.02$. The afterglow
decayed with an index of $-1.36$, consistent with the analysis of
\citet{uehara10}. A flare occurred around $t=500$~s following the
burst and approached the original maximum brightness of $4.23\,{\rm
  \mu Jy}$. After the flare emission became weak, the afterglow
emission again dominated the X-ray light curve and continued to decay
with the same index ($\alpha_{\rm x} = 1.36$) to the end of the XRT
observations. The X-ray flare seems like superpose on the X-ray
decelerated temporal evolution and did not change the overall X-ray
afterglow evolution significantly.

This analysis of X-ray and $R$-band light curve of GRB\,071112C shows
that the X-ray light curve was composed of a single power-law
($\alpha_{\rm x}$ $\sim 1.36$) decay plus a small flare while the
$R-$band light curve exhibits a bump followed by a shallower single
power-law decay ($\alpha_{\rm o}\sim 0.92$).

\section{DISCUSSION}

\subsection{Early Bump in Optical Light Curve}

%An important result in the {\it Swift} era is that many early optical
%observations show that many GRBs display bump-like, plateaus or simple
%power-law decay in their early optical light curves. 

In the {\it Swift} era, many early optical afterglows show localized
peaks, plateaus, or simple power-law decay behavior. A simple power
law decay is usually associated with a relativistic blast wave
decelerated by its interaction with the ambient medium.
Unfortunately, the nature of localized peaks and plateaus are
unclear. \citet{pan08} proposed that the peak and plateau features
could be caused by a structured outflow seen at different directions
from the GRB ejecta. The different off-axis viewing angles produce
different features in early optical light curves. The afterglows with
plateaus have larger viewing angles than those with sharper peaks.

Alternatively, the afterglows with plateaus could be simply produced
by long-lived GRBs which display shallow decay in the light curves and
continue for up to $10^4$~s after the GRB onset
\citep{pan11,kann10}. Besides the interpretations we mentioned above,
the optical afterglow peaks could also be produced by the onset of a
normal afterglow or the passage of the synchrotron typical peak
frequency.
%Alternatively, the optical {\bf afterglow peaks} could {\bf be}
%produced by {\bf release of} impulsive ejecta {\bf [I don't understand
%    what you have in mind here...]}.

Our analysis shows that the GRB\,071112C $R$-band light curve peaked
around 99~s after the burst and then decayed with an index of 0.92
until $6.9\times10^4$~s. The observed temporal evolution is not
consistent with observers located off-axis of the GRB jet
\citep{granot02} which should peak thousands seconds after the initial
occurrence. For the scenario of long-lived GRBs, energy from the GRB
ejecta could continue to supply and power the ambient medium
surrounding the burst. The afterglow emission from the ambient medium,
could continuously be supplied and display plateaus or shallow decays
in afterglow light curves. Similar features should be found in both
the X-ray and optical light curves. Although the rising part in the
$R$-band was not visible in our measurements, the short duration peak
in GRB\,071112C implies that it is unlikely an example of the long
lasting plateau feature of long-lived GRBs \citep{zhang06,pan08}. In
addition, in Figure~2, a comparison of X-ray and $R$-band light curves
of GRB\,071112C, shows very different temporal evolutions for the two
wavelengths. This indicates that the mechanism of long-lasting shallow
decay produced by long-lived GRB ejecta can not explain the observed
X-ray and optical temporal evolution.

%% (2)  Case of onset of afterglow
%% (3)  mu_m pass optical band
For the case of synchrotron frequency passage, the external shock
model \citep{sari98} predicts that the optical temporal light curve
($F_{t,\nu_{\rm opt}} \sim t^{-\alpha} \nu^{-\beta}$) will show an
initial increase with $t^{0.5}$, until the synchrotron peak ($t_{\rm
  m}$) after which a power-law decay $t^{ 3(p-1)/4}$ will follow .
Here $p$ is the electron spectral index and the ambient medium is
assumed uniform. This model predicts that the passing times at
different wavelengths should follow $t_1/t_2 \propto (\nu_1/
\nu_2)^{(-2/3)}$. Chromatic breaks and color change are two
significant clues in multi-band light curves to verify the passage of
the synchrotron peak frequency.
% [Is this what you intended???]}.

Our $R$-band light curve is composed of a possible power law rising
with index $\alpha_{\rm R1} = -1.54 \pm 0.62$, a peak of brightness at
$t_{\rm Rb} = 99.4 \pm 7.3$~s followed by a decay with index
$\alpha_{\rm R2} = 0.92 \pm 0.02$. The $I$-band measurements could not
be well fitted with a single power-law decay, which implies it has
similar temporal property to that observed in the $R$-band. To model
the $I$-band measurements with more complicated formulae, we first
fixed the rising power-law index, $\alpha_{\rm I1} = -1.54$, to be the
same as the $R$-band and then fit $I$-band light curve with a broken
power-law formula. We found that the break time in the $I$-band is
$t_{\rm Ib} = 138.3 \pm 32.7$~s after which follows a power-law decay
$\alpha_{\rm I2} = 1.01 \pm 0.04$ ($\chi^{2}/\nu$= 1.04 for
$\nu=4$). This fit is better than the single power-law fit. If the
$R$-band peak was produced by the passage of the synchrotron peak
frequency, the estimated $I-$band break time from external model will
be at $t_{\rm I} = 115.5 \pm 8.5$~s. The result is consistent with the
break time from a broken power-law fit. However, few $I$-band
measurements yielded large error of $I$-band break time and given the
uncertainties to confirm the synchrotron peak frequency at $I$-band.
Fortunately, there is no significant color change between our $R$- and
$I$-band measurements. An achromatic NIR evolution was also reported
by \citet{uehara10} which supports our optical results and indicates
achromatic evolution in the $R$- and $I$- band light curves. The peak
in the $R$-band is thus unlikely due to the passage of the synchrotron
peak.

In the scenario of onset of afterglow, achromatic evolution is
predicted in the multi-band light curves. Such GRBs are generated
from high relativistic injection fireballs
\citep{meszaros02,zhang04,prian04}. The fireball maintains constant
velocity until it sweeps up a significant amount of ambient medium and
then is decelerated by the ambient medium to produce a smooth local
peak in the afterglow light curve. During this process, the Lorentz
factor, $\Gamma$, decreases. The peak time of the bump, from
theoretical prediction, demonstrates roughly half of the fireball
energy is transferred to the medium and is detectable in the early
afterglow light curve. For some bursts, in which the reverse shock
component would not show up in the optical band, the smooth local peak
signals the deceleration feature of the fireball and can be used to
constrain the initial Lorentz factor and the deceleration radius
\citep{sari99,zhang03,koba07,liang10}. In addition, this theory
predicts that the peak should be sensitive to the initial Lorentz
factor $\Gamma_0$ but is insensitive to other parameters.

\citet{molinari} studied the NIR early peaks of GRB\,060418 and
GRB\,060607A and concluded that such features could be explained by
the onset of afterglows. Their estimated values of initial Lorentz
factor ($\Gamma_0$) are consistent with predictions ($50 \la \Gamma_0
\la 1000$) from the standard fireball model
\citep{piran10,sod02,meszaros06}.  With the formula in
\citet{molinari}, we calculated the expected time that $R$-band light
curve reaches its maximal brightness for GRB\,071112C. The peak time
$t_{\rm peak} = t_{\rm b}(-\alpha_{\rm R1}/\alpha_{\rm
  R2})^{1/({\alpha_{\rm R2}-\alpha_{\rm R1})}}$ is $123 \pm 8$~s. The
initial Lorentz factor $\Gamma_0$ is $\approx 257 \pm 20$ for a
constant density medium and $\approx 69 \pm 6$ for a wind
environment. In the wind environment, the interstellar medium (ISM)
density distribution around a massive star can be defined as $n(r)
= A \times (r)^{-2}$\,${\rm cm}^{-3}$, where $A$ is a
constant. The estimated initial Lorentz factor is consistent with the
theoretical prediction at the lower end. On the other hand,
\citet{liang10} found that when GRBs show onset feature in their early
optical or X-ray light curves, their initial Lorentz factor,
$\Gamma_0$, and GRB isotropic energy, $E_{\rm iso}$, follow an
empirical relation $\Gamma_0 \simeq 182 E_{{\rm iso}, 52}^{0.25 \pm
  0.03}$.  Here $E_{\rm iso,52}$ is the GRB isotropic energy in unit
of $10^{52}$\,erg and a uniform GRB ambient medium is assumed. Our
estimation shows the initial Lorentz factor of GRB\,071112C is
$\Gamma_{0} \sim 260$. Figure~3 depicts the empirical correlation
between ${\rm T}_0$ and $E_{\rm iso, 52}$. With the isotropic energy
at $z = 0.823$, isotropic energy ($E_{\rm iso,52} \sim 0.53$\,erg) of
GRB\,071112C follows the empirical relation within the 2-$\sigma$
range. This analysis thus further supports the conclusion that the
bump in GRB\,071112C is most likely the onset afterglow at optical
wavelengths.

\subsection{Different Origin of X-ray and Optical Emission}

As we have discussed, optical and X-ray light curves in GRB\,071112C
have different evolutions and the rising part of the $R-$band light
curve is likely related to the onset of afterglow. In fact, several
observations and studies show that the X-ray and optical light curves
are often different and inconsistent with the external shock model in
which X-ray and optical emission are produced by the same emission
mechanisms \citep{pan06,urata07,liang09}. Those GRBs usually have
complicated and diverse temporal evolutions.

\citet{urata07} investigated the late temporal properties of 14 GRBs
and found that a fraction of the events are outliers of the external
shock model at normal decay phase in which neither the delayed energy
injection nor time dependency of shock microphysics were
considered. \citet{uehara10} studied the NIR to X-ray spectral energy
distribution of GRB\,071112C and concluded that the observed NIR to
X-ray SEDs is consistent with the expectation from the normal
afterglow component and that the cooling break ($\nu_{\rm c}$) is
between the optical and X-ray bands. In other words, spectral
evolution of observed GRB\,071112C should be in the region of
$\nu_{\rm m} < \nu_{\rm o} < \nu_{\rm c} < \nu_{\rm x}$ and follow the
relationships predicted by external shock models, $\alpha_{\rm o} -
\alpha_{\rm x} > -1/4$ for uniform ambient median or $\alpha_{\rm o} -
\alpha_{\rm x} < 1/4$ for stellar wind with a density variation $\rho
\propto r^{-2}$ \citep{urata07}. The X-ray temporal power-law decay
index, excluding the flare component, is $\alpha_{\rm x} \sim 1.36$
and the optical decay index (after the bump feature) is $\alpha_{\rm
  o} \sim 0.92$. The observed difference of power-law indices,
$\alpha_{\rm o} - \alpha_{\rm x}$ $= -0.44 \pm 0.03 $ for
GRB\,071112C, is outlier of the external shock model and suggests
different origins or radiation processes for X-ray and optical
emissions.

Nearly half of all {\it Swift} bursts have distinct X-ray flares and
they are most likely due to late prompt emission caused by late
central engine activity \citep{zhang06,fal07}. \citet{chin10} and
\citet{bernardini11} investigated early- and late-time X-ray flares
and concluded that the internal shock origin is the most promising
explanation for X-ray flares. Those studies give strongly indications
that X-ray flares have a common origin with the gamma-ray
pulses. Besides, the presence of an underlying continuum with same
slope before and after the flaring activity excluded the possibility
that flares are related to the afterglow emission by forward external
shocks. These investigations implies additional energy is needed to
produce the chromatic temporal properties in X-ray and optical
wavelengths.

To explain the origin of both X-ray and optical evolutions constantly,
\citet{ghise07} proposed that the observed X-ray and optical fluxes
could be modified by two emission components. One is the afterglow
emission produced by forward shocks. Another is late prompt emission,
which has same origin of prompt emission, but is created at late times
with smaller power and smaller $\Gamma$. In this interpretation, if
the X-ray flux is dominated by late prompt emission and the optical
flux is dominated by afterglow emission, the light curves in the two
wavelengths will evolve independently and show no simultaneous break.
A faint X-ray flare found around $t=$500~s following the GRB\,071112C
burst could provide a clue that late prompt emission plays a role in
the X-ray emission.

To further explore late internal dissipation in X-ray emission, we
assume that the observed optical emission in GRB\,071112C is the real
afterglow predicted by the external shock model. In this model (
$\nu_{\rm m} < \nu_{\rm o} < \nu_{\rm c}$$< \nu_{\rm x}$), the optical
emission would follow $F_{t,\nu_{\rm opt}} \sim t^{-\alpha}
\nu^{-\beta} \sim t^{3(p-1)/4} {\nu}^{-(1-p)/2}$ leading to a
relation, $\alpha_{\rm o} = 1.5 \times \beta_{\rm o}$.  With the
observed $R$-band power-law index, $\alpha_{\rm R} = 0.92$, we
estimate spectral index is $\beta_{\rm o} = 0.61$ for the optical band
and calculate the electron spectral index $p = 2.24$. Next we assume
$p$ is constant in afterglow phase and calculate the X-ray spectral
index produced by the external shock model, $\beta{\rm x} = 1.12$. In
Figure~2, the dotted line shows the expected X-ray emission from
external shock model in the region of $\nu_{\rm m} < \nu_{\rm o} <
\nu_{\rm c}$ ($F_{\rm x,exp} \propto$ $F_{\rm opt}$($\nu_{\rm
  x}/\nu_{\rm o})^{-0.61}$). Here we adopt the value $\nu_{\rm c} =
10^7$~Hz from \citet{uehara10}. In addition, we also plot the maximum
value of expected X-ray flux (the dot-dashed line in the
Figure\,2). It is clear that the observed X-ray flux is brighter than
the expected flux from external shock. This supports the conclusion
that the observed X-ray and optical emissions from GRB\,071112C are
caused by different emission mechanisms. The X-ray flux is created by
late internal dissipation and X-ray afterglow emission while the
optical flux is dominated by afterglow. In addition, the expected
X-ray flux in the Figure~2 implies that the late prompt emission could
last until 3000~s after the burst or even longer ($\sim 10^4$~s). This
is consistent with late flares or shallow decay in some bursts, which
are generally believed to be related to the late activity of central
engine \citep{burrows05,nousek06,zhang06,fal06}.

Recently, \citet{pan11} proposed another interpretation. They proposed
that the X-ray and optical evolution could be decoupled by additional
energy added to external shock in a wind-like medium. They suggested
that the optical emission is from synchrotron and the X-ray emission
is from local inverse-Compton scattering. In internal$-$external
model, the fireball ejecta collides with the ambient ISM and produce
synchrotron afterglow emission in X-ray and optical wavelengths. At
this moment, if additional energy is supplied into the ejecta, the low
energy photons from synchrotron processes will obtain energy from
relativistic electrons through inverse-Compton scattering and enhance
the X-ray flux. In this scenario, the X-ray flux is predicted to have
faster decay than the optical flux and no achromatic breaks will be
found in the two wavelengths.
 
\citet{pan11} assumed that an energy injection ($E \sim t^{\rm e}$)
and a power-law distribution of electrons with energy $dN/d\gamma \sim
\gamma^{-p}$ for the synchrotron self-Compton model. With the
conditions, they derived the predicted optical (from synchrotron) and
X-ray (from inverse-Compton) power-law decay indices. For a wind-like
medium, the expected synchrotron decay index is $\alpha_{\rm o} = 1/4
\times[3p-1-(p+1)e]$ at $\nu_m^{\rm sy} < \nu_{\rm o} < \nu_c^{\rm
  sy}$ and the inverse-Compton decay index is $\alpha_{\rm x} =
p-1-(pe/2)$ at $\nu_c^{\rm ic} < \nu_{\rm x}$. Using our results on
GRB\,071112C ($\alpha_{\rm o} = 0.92$ and $\alpha_{\rm x} = 1.36$), we
derived a relation between energy injection and electron spectral
index $p = e -0.04$. Applying the electron spectral index $p=2.24$
from optical observation, we obtain $e$ = 2.2 for the
GRB\,071112C. This result is consistent with other afterglows in which
their decoupled X-ray and optical light curves can be explained by
synchrotron self-Compton model (GRB\,080129 with $e \simeq$ 2.0,
GRB\,090424 with $e \simeq $ 1.0, and GRB\,090510 with $e \simeq$
2.4).

We investigated the X-ray and optical temporal evolution of the
GRB\,071112C. Our analysis shows that different emission mechanisms
produce the decoupled X-ray and optical evolution. The optical flux is
dominated by afterglow, which is produced by synchrotron
emission. However, the X-ray flux is created by an additional feature
related to energy injection and X-ray afterglow emission. Different
theoretical interpretations indicate the additional feature in X-ray
can be explained by either late internal dissipation or
inverse-Compton scattering in external shocks.

\section{Conclusion}
We analyzed X-ray and optical light curves of GRB\,071112C and found
that the X-ray light curve was described by a single power-law plus a
flare-like feature, while the $R$-band light curve showed an initial
rise followed by a power-law decay. No significant color changes and a
value of $\Gamma_0 = 257 \pm 20$ for initial Lorentz factor
indicates that the afterglow onset scenario is likely the correct
interpretation for the early $R$-band rise.  Based on the result, we
conclude that the optical flux of the GRB\,071112C is dominated by
afterglow. Furthermore, compared with X-ray temporal evolution, we
found that the observed temporal properties in the two wavelengths
cannot be described by the external shock in which the X-ray and
optical emission are produced by the same emission mechanism. An
additional energy contribution in X-rays is thus necessary. The X-ray
flux could be created by a additional feature related to energy
injection and X-ray afterglow emission. The faint X-ray flare supports
the scenario of energy injection and our analysis indicates either
late internal dissipation or inverse-Compton scattering in external
shocks is the possible interpretation for the additional feature by
energy injection.  More such samples with adequately sampled X-ray and
optical light curves are important to investigate and understand the
detailed emission mechanism for the two wavelengths.

\acknowledgments 

This work is supported by grants NSC-99-2112-M-001-002-MY3 (K.Y.H.) and
NSC-99-2112-M-008-003-MY3 (Y.U.). The TAOS project is supported in part
by the thematic research program AS-88-TP-A02 at Academia Sinica and
the ROTSE project is supported by NASA Grant NNX08AV63G and NSF Grant
PHY-0801007. Access to the CFHT was made possible by the Ministry of
Education and the National Science Council of Taiwan as part of the
Cosmology and Particle Astrophysics (CosPA) initiative. This work made
use of data supplied by the UK Swift Science Data Center at the
University of Leicester.


%% in parentheses, after the keyword, but they will not be verified.
{\it Facilities:} \facility{Swift(XRT)}, \facility{CFHT}



\appendix




\clearpage
%=======
% FIg 1
%=======

\begin{figure}
\includegraphics[angle=270,scale=0.6]{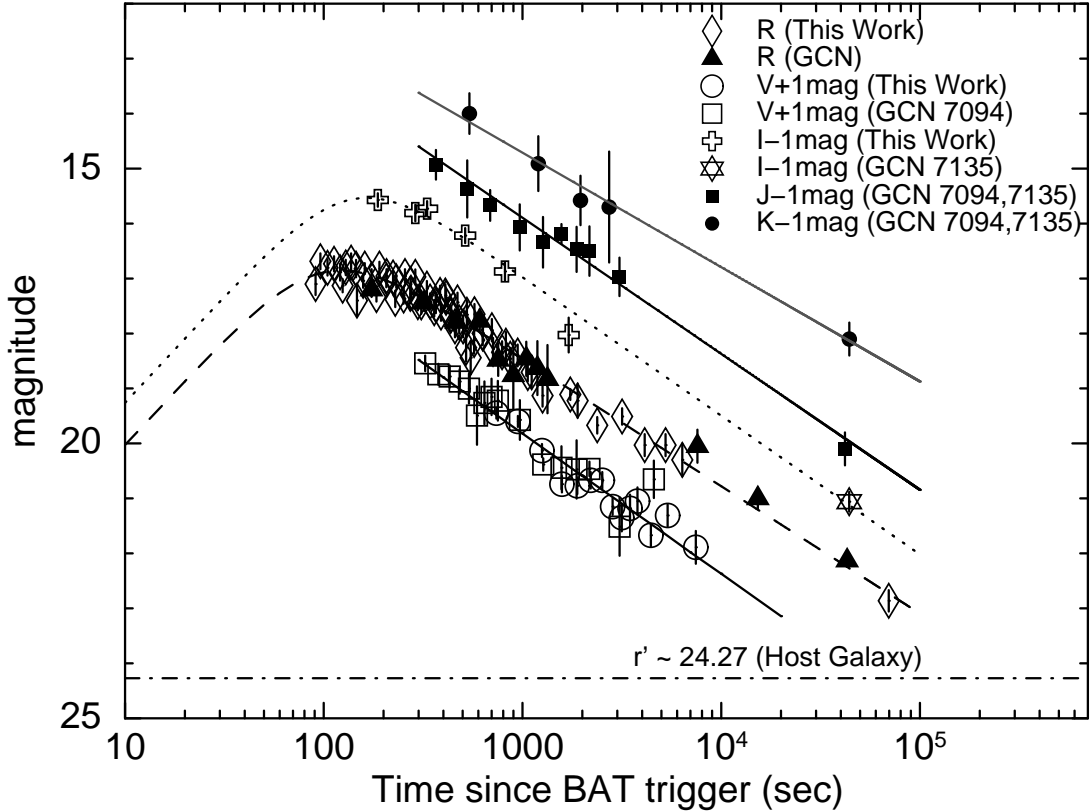}
\rotate
\caption{Optical light curves of GRB\,071112C: the solid line presents
  the best fit by the single power-law model [$F \propto t^{-\alpha}$]
  for the $V$-band ($\alpha_{\rm V} = 1.02 \pm 0.05$), $J$-band
  ($\alpha_{\rm V} = 0.99 \pm 0.04$), and the $K$-band ($\alpha_{\rm
    V} = 0.83 \pm 0.04$). The dashed and dotted lines indicate the
  best fit by the broken power law [$F(\nu,t) = {F_{\nu}^{\ast}} /
    [(t/t_{\rm b})^{\alpha_1}+(t/t_{\rm b})^{\alpha_2}]$] with the
  $R$-band ($\alpha_{\rm R1} = -1.54 \pm 0.62$, $\alpha_{\rm R2} =
  0.92 \pm 0.02$,and $t_{\rm Rb} = 99.4 \pm 7.3$~s) and with the
  $I$-band ($\alpha_{\rm I1} = -1.54$ (fix), $\alpha_{\rm I2} = 1.01
  \pm 0.04$,and $t_{\rm Ib} = 138.3 \pm 32.7$~s), respectively. The
  dot-dashed line shows $r'$-band brightness of GRB\,071112C host
  galaxy.}
\label{index_ox}
\end{figure}
\clearpage

%=======
% FIg 2
%=======
\begin{figure}
\includegraphics[angle=270,scale=0.6]{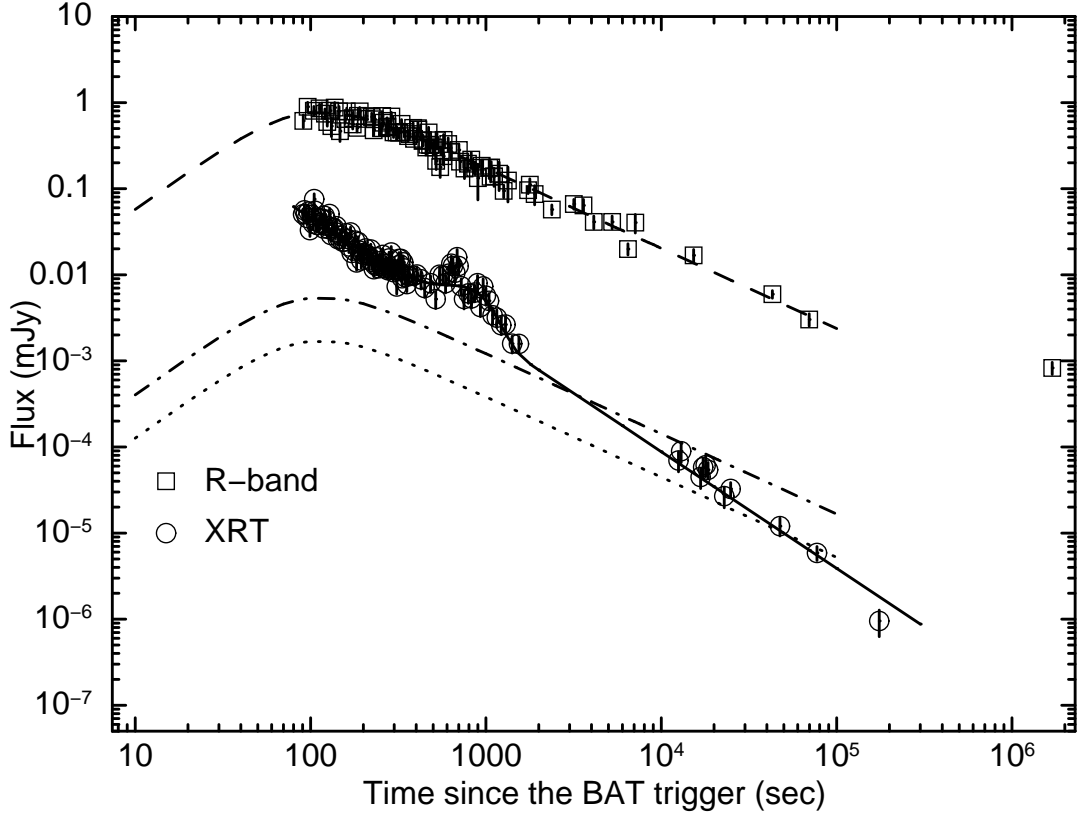}
\caption{Observed X-ray and the $R$-band light curves and the expected
  X-ray flux from external shock model: the dashed line shows the best
  $R$-band fit presented in the Figure~1. The solid line presents the
  best fit with X-ray by a single power-law decay plus a Gaussian
  function. The dot-dashed line presents the maximum expected X-ray
  flux from external shock model($F_{\rm x,exp} \propto$ $F_{\rm
    opt}$($\nu_{\rm x}/\nu_{\rm o})^{-0.61}$). The dot line presents
  the expected X-ray flux in the region of $\nu_{\rm o} < \nu_{\rm c}
  < \nu_{\rm x}$ ($F_{\rm x,exp} \propto$ $F_{\rm opt}$ ($\nu_{\rm
    c}/\nu_{\rm o})^{-0.61}$($\nu_{\rm x}/\nu_{\rm c})^{-1.12}$), here
  the value $\nu_{\rm c} = 10^7$~Hz is from \citet{uehara10}.}
\label{xrayflux}
\end{figure}
\clearpage

%=======
% FIg 3
%=======
\begin{figure}
\includegraphics[angle=270,scale=0.6]{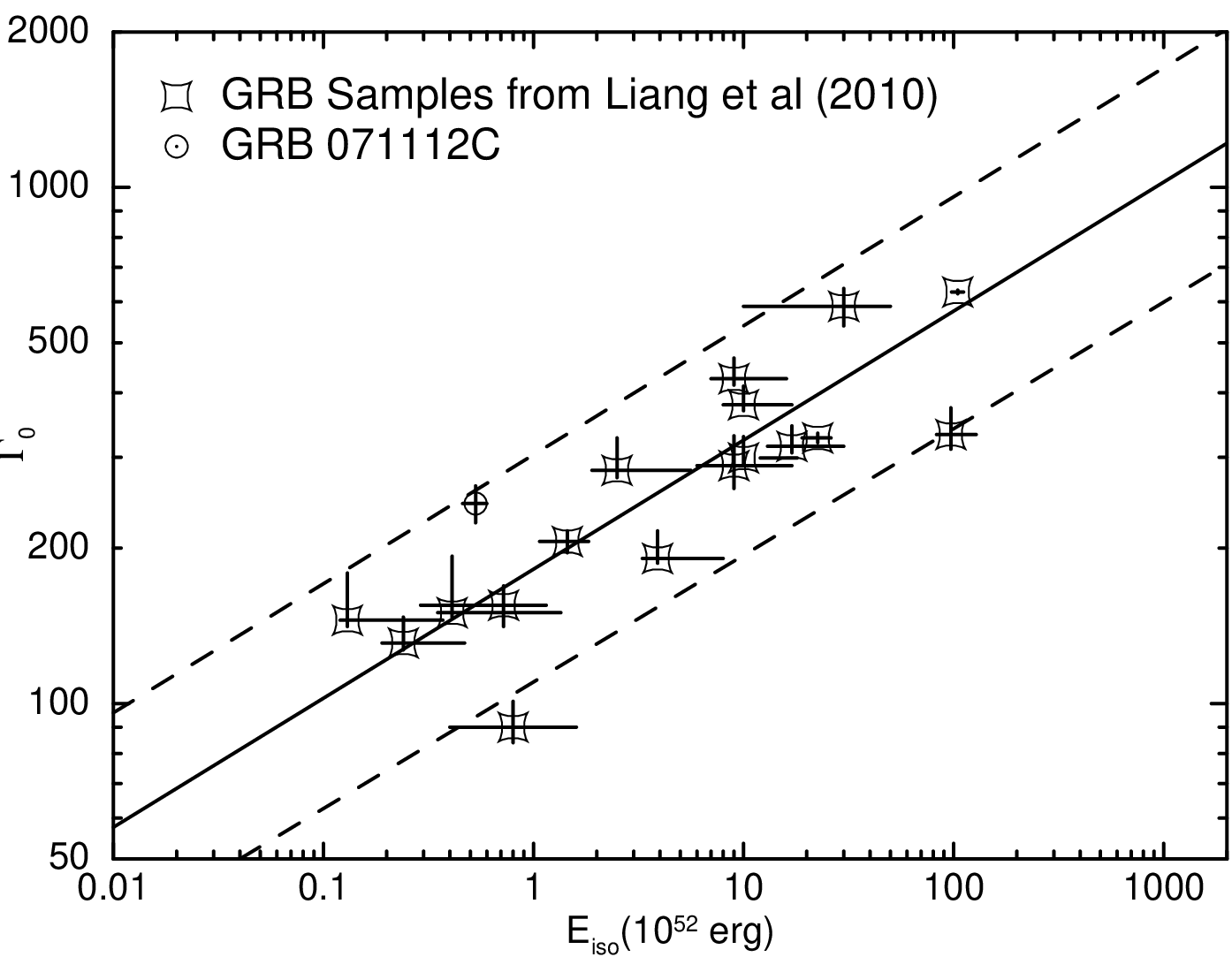}
\caption{Correlation between $\Gamma_0$ and $E_{\rm iso}$. The open
  stars shows optical selected samples from \citet{liang09} and the
  open circle presents the value of GRB\,071112C derived from our
  $R$-band measurements. The solid and dashed lines are $\Gamma_0
  \simeq 182 E_{\gamma, {\rm iso}, 52}^{0.25 \pm 0.03}$ and
  2\,$\sigma$, respectively \citep{liang09}. This diagram shows that
  the derived initial Lorentz factor ($\Gamma_0$) and isotropic energy
  are fit in with other GRBs with bump-like feature in early optical
  light curves. }
\label{initial}
\end{figure}

\clearpage
%==========
% Table 1
%==========

\begin{deluxetable}{llccc}
\tablecaption{Observation Log of GRB\,071112C Optical Afterglow}
\tablewidth{0pt}
\tablehead{
  \colhead{$T_{\rm mid}({\rm s})$}& \colhead{Filter}  & 
  \colhead{Exposure (s)} & \colhead{Mag} & \colhead{Telescope} }
\startdata
%=========== Xinglong V ==============================================
738.7 & $V$ & 300~s $\times$ 1 & 18.45 $\pm$ 0.10 & {\rm Xinglong-1m} \\
948.9 & $V$ & 300~s $\times$ 1 & 18.60 $\pm$ 0.10 & {\rm Xinglong-1m}\\
1252.8 & $V$ & 300~s $\times$ 1 & 19.13 $\pm$ 0.12 & {\rm Xinglong-1m} \\
1578.5 & $V$ & 300~s $\times$ 1 & 19.74 $\pm$ 0.15 & {\rm Xinglong-1m}\\
1887.8 & $V$ & 300~s $\times$ 1 & 19.77 $\pm$ 0.15 & {\rm Xinglong-1m}\\
2204.9 & $V$ & 300~s $\times$ 1 & 19.67 $\pm$ 0.15 & {\rm Xinglong-1m}\\
2522.0 & $V$ & 300~s $\times$ 1 & 19.67 $\pm$ 0.15 & {\rm Xinglong-1m} \\
2839.1 & $V$ & 300~s $\times$ 1 & 20.15 $\pm$ 0.15 & {\rm Xinglong-1m} \\
3157.1 & $V$ & 300~s $\times$ 1 & 20.35 $\pm$ 0.25 & {\rm Xinglong-1m} \\
3474.1 & $V$ & 300~s $\times$ 1 & 20.18 $\pm$ 0.25 & {\rm Xinglong-1m}\\
3791.2 & $V$ & 300~s $\times$ 1 & 20.05 $\pm$ 0.25 & {\rm Xinglong-1m} \\
4425.4 & $V$ & 300~s $\times$ 3 & 20.67 $\pm$ 0.20 & {\rm Xinglong-1m}\\
5377.5 & $V$ & 300~s $\times$ 3 & 20.31 $\pm$ 0.20 & {\rm Xinglong-1m} \\
7454.6 & $V$ & 600~s $\times$ 5 & 20.89 $\pm$ 0.30 & {\rm Xinglong-1m}\\
%========== ROTSE =================================================
 90.9  & $CR$ & 5~s $\times$ 5  & 17.10 $\pm$ 0.20 & {\rm ROTSE-IIIc} \\
147.0  & $CR$ & 5~s $\times$ 5 & 17.40 $\pm$ 0.30 & {\rm ROTSE-IIIc} \\
273.3  & $CR$ & 20~s $\times$ 5 & 17.10 $\pm$ 0.10 & {\rm ROTSE-IIIc} \\
% ========= TAOS ==================================================
 96.2 & $R$ & 5~s $\times$ 1 & 16.79 $\pm$ 0.15 & {\rm TAOSB}       \\
104.4 & $R$ & 5~s $\times$ 1 & 17.01 $\pm$ 0.17 & {\rm TAOSB}       \\
112.6 &$R$ & 5~s $\times$ 1 & 16.83 $\pm$ 0.14 & {\rm TAOSB}       \\
120.8 &$R$ & 5~s $\times$ 1 & 17.05 $\pm$ 0.15 & {\rm TAOSB}       \\
129.0 &$R$ & 5~s $\times$ 1 & 17.00 $\pm$ 0.16 & {\rm TAOSB}       \\
137.2 &$R$ & 5~s $\times$ 1 & 16.86 $\pm$ 0.17 & {\rm TAOSB}       \\
145.4 &$R$ & 5~s $\times$ 1 & 16.96 $\pm$ 0.16 & {\rm TAOSB}       \\
157.7 &$R$ & 5~s $\times$ 2 & 17.17 $\pm$ 0.12 & {\rm TAOSB}       \\
174.1 &$R$ & 5~s $\times$ 2 & 17.17 $\pm$ 0.18 & {\rm TAOSB}       \\
190.5 &$R$ & 5~s $\times$ 2 & 16.96 $\pm$ 0.14 & {\rm TAOSB}       \\
206.9 &$R$ & 5~s $\times$ 2 & 17.11 $\pm$ 0.15 & {\rm TAOSB}       \\
223.3 &$R$ & 5~s $\times$ 2 & 17.14 $\pm$ 0.12 & {\rm TAOSB}       \\
231.7 &$R$ & 5~s $\times$ 2 & 17.16 $\pm$ 0.12 & {\rm TAOSB}       \\
256.1 &$R$ & 5~s $\times$ 2 & 17.11 $\pm$ 0.14 & {\rm TAOSB}        \\
272.5 & $R$ & 5~s $\times$ 2 & 17.42 $\pm$ 0.18 & {\rm TAOSB}       \\
288.9 & $R$ & 5~s $\times$ 2 & 17.14 $\pm$ 0.16 & {\rm TAOSB}       \\
309.5 & $R$ & 5~s $\times$ 3 & 17.52 $\pm$ 0.22 & {\rm TAOSB}        \\
330.3 & $R$ & 5~s $\times$ 3 & 17.28 $\pm$ 0.13 & {\rm TAOSB}       \\
359.7 & $R$ & 5~s $\times$ 3 & 17.63 $\pm$ 0.15 & {\rm TAOSB}       \\
384.6 & $R$ & 5~s $\times$ 3 & 17.48 $\pm$ 0.17 & {\rm TAOSB}       \\
409.2 & $R$ & 5~s $\times$ 3 & 17.45 $\pm$ 0.22 & {\rm TAOSB}       \\
437.9 & $R$ & 5~s $\times$ 4 & 17.75 $\pm$ 0.18 & {\rm TAOSB}       \\
470.7 & $R$ & 5~s $\times$ 4 & 17.57 $\pm$ 0.15 & {\rm TAOSB}       \\
520.3 & $R$ & 5~s $\times$ 4 & 18.42 $\pm$ 0.31 & {\rm TAOSB}       \\
573.2 & $R$ & 5~s $\times$ 7 & 17.81 $\pm$ 0.19 & {\rm TAOSB}       \\
630.7 & $R$ & 5~s $\times$ 7 & 18.14 $\pm$ 0.24 & {\rm TAOSB}       \\
701.1 & $R$ & 5~s $\times$ 7 & 18.15 $\pm$ 0.22 & {\rm TAOSB}       \\
824.7 & $R$ & 5~s $\times$ 20 & 18.36 $\pm$ 0.22 & {\rm TAOSB}      \\
1112.4 & $R$ & 5~s $\times$ 50 & 19.01 $\pm$ 0.29 & {\rm TAOSB}      \\
1894.6 & $R$ & 5~s $\times$ 140 & 19.23 $\pm$ 0.26 & {\rm TAOSB}     \\
%=============== Xinglong 0.8m ========================================
124.4 & $CR$ & 20~s $\times$ 1 & 17.13 $\pm$ 0.12 & {\rm TNT-0.8m}\\
160.7 & $CR$ & 20~s $\times$ 1 & 16.82 $\pm$ 0.10 & {\rm TNT-0.8m}\\
184.0 & $CR$ & 20~s $\times$ 1 & 17.30 $\pm$ 0.12 & {\rm TNT-0.8m}\\
206.5 & $CR$ & 20~s $\times$ 1 & 17.04 $\pm$ 0.12 & {\rm TNT-0.8m}\\
228.9 & $CR$ & 20~s $\times$ 1 & 17.37 $\pm$ 0.15 & {\rm TNT-0.8m} \\
252.3 & $CR$ & 20~s $\times$ 1 & 17.31 $\pm$ 0.15 & {\rm TNT-0.8m}\\
274.8 & $CR$ & 20~s $\times$ 1 & 17.26 $\pm$ 0.15 & {\rm TNT-0.8m} \\
297.2 & $CR$ & 20~s $\times$ 1 & 17.43 $\pm$ 0.15 & {\rm TNT-0.8m}\\
343.0 & $CR$ & 20~s $\times$ 1 & 17.42 $\pm$ 0.15 & {\rm TNT-0.8m} \\
366.3 & $CR$ & 20~s $\times$ 1 & 17.48 $\pm$ 0.15 & {\rm TNT-0.8m}\\
 388.8 & $CR$ & 20~s $\times$ 1 & 17.62 $\pm$ 0.15 & {\rm TNT-0.8m}   \\
412.1 & $CR$ & 20~s $\times$ 1 & 17.35 $\pm$ 0.15 & {\rm TNT-0.8m}\\
434.6 & $CR$ & 20~s $\times$ 1 & 17.68 $\pm$ 0.15 & {\rm TNT-0.8m} \\
457.0 & $CR$ & 20~s $\times$ 1 & 17.87 $\pm$ 0.20 & {\rm TNT-0.8m}\\
480.4 & $CR$ & 20~s $\times$ 1 & 17.80 $\pm$ 0.20 & {\rm TNT-0.8m} \\
502.8 & $CR$ & 20~s $\times$ 1 & 17.87 $\pm$ 0.20 & {\rm TNT-0.8m}\\
525.3 & $CR$ & 20~s $\times$ 1 & 17.66 $\pm$ 0.25 & {\rm TNT-0.8m} \\
548.6& $CR$ & 20~s $\times$ 1 & 18.44 $\pm$ 0.30 & {\rm TNT-0.8m} \\
571.1 & $CR$ & 20~s $\times$ 1 & 18.12 $\pm$ 0.25 & {\rm TNT-0.8m} \\
632.4 & $R$ & 60~s $\times$ 1 & 18.00 $\pm$ 0.15 & {\rm TNT-0.8m}    \\
711.1 & $R$ & 60~s $\times$ 1 & 18.28 $\pm$ 0.20 & {\rm TNT-0.8m}    \\
789.7 & $R$ & 60~s $\times$ 1 & 18.34 $\pm$ 0.20 & {\rm TNT-0.8m}    \\
867.5 & $R$ & 60~s $\times$ 1 & 18.46 $\pm$ 0.20 & {\rm TNT-0.8m}    \\
946.2 & $R$ & 60~s $\times$ 1 & 18.40 $\pm$ 0.20 & {\rm TNT-0.8m}     \\
1063.6 & $R$ & 60~s $\times$ 2 & 18.72 $\pm$ 0.15 & {\rm TNT-0.8m}    \\
1262.3 & $R$ & 60~s $\times$ 3 & 19.13 $\pm$ 0.20 & {\rm TNT-0.8m}     \\
1740.1 & $R$ & 300~s $\times$ 2 & 19.11 $\pm$ 0.10 & {\rm TNT-0.8m}    \\
2376.0  & $R$ & 300~s $\times$ 2 & 19.67 $\pm$ 0.15 & {\rm TNT-0.8m}      \\
3170.0 & $R$ & 300~s $\times$ 3 & 19.51 $\pm$ 0.15 & {\rm TNT-0.8m}   \\
4123.0 & $R$ & 300~s $\times$ 3 & 20.03 $\pm$ 0.20 & {\rm TNT-0.8m}   \\
5243.6 & $R$ & 300~s $\times$ 4 & 20.03 $\pm$ 0.20 & {\rm TNT-0.8m}   \\
186.0 & $Rc$ & 60~s $\times$ 1 & 17.11 $\pm$ 0.11 & {\rm MITSuMe}   \\
280.5 & $Rc$ & 60~s $\times$ 2 & 17.46 $\pm$ 0.11 & {\rm MITSuMe}   \\
536.5 & $Rc$ & 60~s $\times$ 5 & 18.03 $\pm$ 0.15 & {\rm MITSuMe}   \\
911.5 & $Rc$ & 60~s $\times$ 5 & 18.49 $\pm$ 0.23 & {\rm MITSuMe}    \\
1934.5 & $Rc$ & 60~s $\times$ 19 & 19.21 $\pm$ 0.25 & {\rm MITSuMe}   \\
6372.0 & $Rc$ & 300~s $\times$ 6 & 20.3 $\pm$ 0.3 & {\rm LOT}         \\
69620.3 & $Rc$ & 300~s $\times$ 12 & 22.7 $\pm$ 0.2 & {\rm LOT}       \\
3.3$\times10^7$ & $r'$ & 300~s $\times$ 4 & 24.27 $\pm$ 0.19 & {\rm CFHT}     \\
187.0& $Ic$ & 60~s $\times$ 1 & 16.57 $\pm$ 0.12 & {\rm MITsuMe}       \\
289.0 & $Ic$ & 60~s $\times$ 1 & 16.81 $\pm$ 0.15 & {\rm MITSuMe}       \\
331.0 & $Ic$ & 60~s $\times$ 1 & 16.73 $\pm$ 0.14 & {\rm MITSuMe}       \\
514.5 & $Ic$ & 60~s $\times$ 2 & 17.22 $\pm$ 0.14 & {\rm MITSuMe}     \\
816.0 & $Ic$ & 60~s $\times$ 6 & 17.87 $\pm$ 0.17 & {\rm MITSuMe}       \\
1707.5 & $Ic$ & 60~s $\times$ 6 & 19.03 $\pm$ 0.32 & {\rm MITSuMe}    \\
%================= 
\enddata
\end{deluxetable}

\end{document}